\documentstyle[aps,preprint]{revtex}

\def\be{\begin{equation}}
\def\ee{\end{equation}}
\def\ba{\begin{eqnarray}}
\def\ea{\end{eqnarray}}

\begin{document}
\draft
\preprint{Fermilab-Pub-96/091-A, astro-ph/9605???}
\date{May 22, 1996}
\title{Gravitational lensing of gravitational waves \\
from merging neutron star binaries}

\author{Yun Wang$^{1}$, Albert Stebbins$^{1}$, and Edwin L. Turner$^{2}$}
\address{
$^{1}$NASA/Fermilab Astrophysics Center, FNAL, Batavia, IL~~60510}

\address{
$^{2}$ Princeton University Observatory, Peyton Hall, Princeton, NJ~~08544}

\maketitle

\begin{abstract}

We discuss the gravitational lensing of gravitational waves from merging
neutron star binaries, in the context of advanced LIGO type gravitational wave
detectors.  We consider properties of the expected observational data with cut
on the signal-to-noise ratio $\rho$, i.e., $\rho>\rho_0$.  An advanced LIGO
should see unlensed inspiral events with a redshift distribution with cut-off
at a redshift $z_{\rm max} < 1$ for $h \leq 0.8$.  Any inspiral events detected
at $z>z_{\rm max}$ should be lensed.  We compute the expected total number of
events which are present due to gravitational lensing and their redshift 
distribution for an advanced LIGO in a flat Universe.  If the matter 
fraction in compact lenses is close to 10\%, an
advanced LIGO should see a few strongly lensed events per year with $\rho >5$.

\end{abstract}

\pacs{PACS numbers: 98.62.Sb, 98.80.Es, 04.80.Nn }

\newpage

\narrowtext

\section{Introduction}

An advanced LIGO may observe gravitational waves produced as distant close
neutron star binary pairs spiral into each other.  During the last stage of
inspiral the binary emits copious gravitational waves, with increasing
frequency as the orbital period decreases, until finally the pair collides and
coalesces. LIGO aims to detect the waves emitted during the last 15 minutes of
inspiral when the frequency sweeps up from 10 Hz to approximately $10^3$ Hz
\cite{Thorne95}.

In this paper, we discuss the gravitational lensing of gravitational waves from
merging neutron star binaries, in the context of advanced LIGO type
gravitational wave detectors.  Following Ref.{\cite{Finn96}}, we consider
properties of the expected observational data with cut on the signal-to-noise
ratio $\rho$, i.e., $\rho>\rho_0$.  An advanced LIGO should see unlensed events
with a redshift distribution with cut-off at a small redshift $z_{\rm max} < 1$
for $h \leq 0.8$ \cite{Finn96,Wang96}. We argue below that there may be a
significant number of inspiral events detected at $z>z_{\rm max}$ which can be
detected because they are magnified due to gravitational lensing.  We compute
the expected total number of events 
which are present due to gravitational lensing
and their redshift distribution for
an advanced LIGO, for plausible choices of cosmological parameters.

The aforementioned frequency range over which LIGO can detect neutron star
binary inspirals corresponds to a wavelength of gravitational waves 
from $3\times 10^4$ to $10^2$ km.  This wavelength will be much 
smaller than the characteristic scales of gravitational fields the 
gravitational waves are likely to encounter as they pass between the 
neutron star binaries and the Earth.  This means that one may
treat the propagation of gravitational waves in the geometrical optics
limit \cite{Book73}.  In other words the gravitational lensing magnification
will be the same as for optical light and one may use the standard formulae
from optical gravitational lens theory.

\section{Observation of unlensed events}

Neutron star binary merger rate at redshift $z$ per unit observer 
time interval per unit volume is 
$\dot{n}_m = \dot{n}_0 \, (1+z)^2 \, \eta(z)$,
where $\dot{n}_0$ is the local neutron star binary merger rate per 
unit volume,
$(1+z)^2$ accounts for the shrinking of volumes with redshift (assuming 
constant comoving volume density of the merger rate) and time dilation, 
and $\eta(z)=(1+z)^{\beta}$ describes evolutionary effects.
We use the ``best guess'' local rate density, 
$\dot{n}_0 \simeq \left(9.9+ 0.6\, h^2\right)\, h \times 10^{-8} 
{\rm Mpc}^{-3} {\rm yr}^{-1} \simeq 10^{-7} h\, {\rm Mpc}^{-3} {\rm yr}^{-1}$.
\cite{Phinney91,NaraPirShe91}. 

In the last stage of a neutron star binary inspiral, gravitational 
radiation energy losses should lead to highly circular binary orbits.
In the Newtonian/quadrupole approximation, for a circular orbit,
the rate at which the frequency of the gravitational waves sweeps 
up or ``chirps'', is determined 
solely by the binary's ``intrinsic chirp mass'',
${\cal M}_0 \equiv (M_1 M_2)^{3/5}/(M_1+M_2)^{1/5}$,
where $M_1$ and $M_2$ are the two bodies' masses.
For a binary inspiral source located at redshift $z$,
the detectors measure ${\cal M} \equiv {\cal M}_0 (1+z)$, 
which is referred to as the {\it observed} chirp mass. 
For a given detector, the signal-to-noise ratio is \cite{Finn96}
\begin{equation}
\rho(z)= 8 \Theta\, \frac{r_0}{d_L(z)}\left( \frac{ {\cal M}(z)}
{1.2 {\rm M}_{\odot}}\right)^{5/6} \, \zeta(f_{\rm max}),
\label{eq:rho}
\end{equation}
$d_L$ is our luminosity distance to the binary inspiral source.
$r_0$ and $\zeta(f_{\rm max})$ depend only on the detector's
noise power spectrum.  
The characteristic distance $r_0$ gives an overall sense of the depth
to which the detector can ``see''. For advanced LIGO, $r_0=355\,$Mpc. 
$\zeta(f_{\rm max})$ reflects the overlap of the signal power 
with the detector bandwidth ($0 \leq \zeta \leq 1$).
For source redshift $z$, $\zeta \simeq 1$ for $1+z \leq 10\,[2.8 
{\rm M}_{\odot}/(M_1+M_2)]$. $\zeta \simeq 1$ is 
a good approximation in the context of this paper.
$\Theta$ is the angular orientation function, it arises from
the dependence of $\rho$ on the relative orientation of the source
and the detector, $0\leq \Theta \leq 4$. Although $\Theta$ can not
be measured, its probability distribution has been found numerically
in Ref.\cite{FinnCher93},
$P_{\Theta}(\Theta, 0\leq \Theta\leq 4)\simeq 5\Theta (4-\Theta)^3 /256$,
$P_{\Theta}(\Theta, \Theta> 4)=0$.

The luminosity distance $d_L(z)=(1+z)^2 d_A(z)$, where $d_A(z)$
is the angular diameter distance. In a flat Universe with a
cosmological constant $\Omega_{\Lambda} =1-\Omega_0 \geq 0$, \cite{Lambda}
$d_A(z)= c H_0^{-1}(1+z)^{-1}\,\int^z_0  {\rm d} w \,
\left[ \Omega_0 \,(1+w)^3 + \Omega_{\Lambda} \right]^{-1/2}.$

The number rate of binary inspiral events seen by a detector on Earth 
with signal-to-noise ratio $\rho>\rho_0$
per source redshift interval is \cite{Finn96,Wang96}
\begin{equation}
\label{eq:NL-P(z)}
\frac{{\rm d}\dot{N}_{NL}(>\rho_0)}{{\rm d} z}=
4\pi \dot{n}_0 \left(cH_0^{-1}\right)^3 
\, \left[ \frac{d_A(z)}{cH_0^{-1}}\right]^2\,\frac{ (1+z)\,\eta(z)}
{ \sqrt{ \Omega_0\, (1+z)^3 + \Omega_{\Lambda}}}\,C_{\Theta}(x),
\end{equation}
where $C_{\Theta}(x)\equiv \int^{\infty}_x {\rm d}\Theta\, 
P_{\Theta}(\Theta)$ is the probability that a given detector 
detects a binary inspiral at redshift $z$ with signal-to-noise 
ratio greater than $\rho_0$, it decreases with $z$ and acts as a 
window function; 
$C_{\Theta}(x, 0\leq x\leq 4)=(1+x)\,(4-x)^4 /256$, 
$C_{\Theta}(x, x>4)=0$.
$x$ is the minimum angular orientation function 
\be
\label{eq:x}
x =\frac{4}{h\, A(r_0, \rho_0, {\cal M}_0)}\,
(1+z)^{7/6}\left[\frac{d_A(z)}{cH_0^{-1}}\right], 
\ee
where we have defined parameter $A$ as in Ref.{\cite{Wang96}},
\begin{equation}
A\equiv 0.4733\,\left(\frac{8}{\rho_0}\right)\, \left(\frac{r_0}{355 \,
\mbox{Mpc}}\right)\, \left( \frac{ {\cal M}_0}{1.2 
{\rm M}_{\odot}}\right)^{5/6}.
\label{eq:A}
\end{equation}
Note that $A$ absorbs the dependence on detector and source properties
$\rho_0$, $r_0$, and ${\cal M}_0$.

Given the detector threshold in terms of the minimum signal-to-noise ratio 
$\rho_0$, the maximum redshift of the source that the detector can ``see'',
$z_{\rm max}$, is given by Eq.(\ref{eq:rho}) with $\Theta=4$.
For advanced LIGO, $z_{\rm max}<1$ for $h \leq 0.8$.
\cite{Finn96,Wang96}
The redshift distribution given by Eq.(\ref{eq:NL-P(z)}) terminates
at $z=z_{\rm max}$.

\section{Observation of lensed events}

For a source at redshift $z^*>z_{\rm max}$, we denote its signal-to-noise
ratio without lensing by $\rho(z^*)$ [given by Eq.(\ref{eq:rho})] . 
The source can not be detected in the absence of lensing.
With lensing, its signal-to-noise ratio becomes
\be
\rho^*(z^*) = \sqrt{\mu}\, \rho(z^*),
\ee  
where $\mu$ is the magnification.
The source can be detected if $\rho^*(z^*)>\rho_0$,
with $\rho_0$ denoting the detector threshold.

The probability of a source at redshift $z$ being magnified by a factor
greater than $\mu$ is $P(>\mu, z)= \tau_L(z)\, y^2(\mu)$, for 
$\tau_L(z) \ll 1$. $\tau_L(z)$ is the optical depth for gravitational 
lensing, and $y^2(\mu) \simeq \mu^{-2}$ for $\mu \gg 1$.
For point mass lenses, we use
\be
y^2(\mu)=\left\{ \begin{array}{ll}
& 2\, \left( \frac{\mu}{\sqrt{\mu^2-1}}-1\right), 
\hskip 1.5cm \mu> \frac{3}{\sqrt{5}} \\
& 1, \hskip 1.5cm \mu \leq \frac{3}{\sqrt{5}}. \end{array}\right.
\ee
The above equation leads to underestimation of $y^2(\mu)$
for $\mu$ close to 1, which has negligible effect for our purpose.
For a source at redshift $z^*>z_{\rm max}$ to be detected, 
we need $\mu>\mu_0$, with
\be
\mu_0 \equiv \left[ \frac{\rho_0}{\rho(z^*)} \right]^2
=\left[\frac{x(z^*)}{\Theta}\right]^2,
\ee
where $x(z^*)$ is given by Eq.(\ref{eq:x}).
Note that $\mu_0$ depends on the angular orientation function $\Theta$.
The larger $\Theta$, the larger the signal-to-noise ratio without lensing
[see Eq.(\ref{eq:rho})], the smaller the magnification needed to
reach the detector threshold $\rho_0$.
Since $\Theta=x$ is the minimum angular orientation function needed for
a source to be seen without lensing [see Eq.(\ref{eq:x})], 
only events with $\Theta<x$ need be considered when we count the number of
events which are present due to gravitational lensing.

The number rate of binary inspiral events which can be seen due to
gravitational lensing by a detector 
on Earth with signal-to-noise ratio $\rho>\rho_0$
per source redshift interval is
\be
\label{eq:L-P(z):0}
\frac{{\rm d}\dot{N}_{L}(>\rho_0)}{{\rm d} z^*} =
4\pi \dot{n}_0 \left(cH_0^{-1}\right)^3 
\, \left[ \frac{d_A(z^*)}{cH_0^{-1}}\right]^2\,\frac{ (1+z^*)\,
\eta(z^*) \, \tau_L(z^*)}
{ \sqrt{ \Omega_0\, (1+z^*)^3 + \Omega_{\Lambda}}}
\int_0^{x(z^*)} {\rm d}\Theta \,P_{\Theta}(\Theta) \,y^2(\mu_0).
\ee
Note that $\Theta_{\rm min}=0$,  because we take the maximum magnification 
to be infinite, which is a reasonable approximation in the context of 
this paper.

We consider two types of lensing: 1) ``macro-lensing'' from the large-scale
gravitational field of galaxies, and 2) ``micro-lensing'' from the smaller
scale gravitational field from compact objects such as stars.  The optical
depth from macrolensing is \cite{Turner90}
\be
\tau_L^G(z) = \frac{F}{30} \, \left[ \frac{ (1+z)\, d_A(z)}{c H_0^{-1}}
\right]^3, 
\ee
where $F$ parametrizes the gravitational lensing effectiveness of
galaxies [as singular isothermal spheres].  Denoting the matter fraction in
compact lenses as $\Omega_{\rm L}$, the optical depth of microlensing is
\cite{TOG,FT91}
\be
\tau_L^p(z) =\frac{3}{2}\,
\frac{\Omega_{\rm L}}{\lambda(z)}
\int^z_0 {\rm d}w\, \frac{(1+w)^3\, \left[\lambda(z)-\lambda(w)\right]\,\lambda(w)}
{\sqrt{\Omega_0(1+w)^3+\Omega_{\Lambda}}},
\ee
where the affine distance (in units of $cH_0^{-1}$) is
$\lambda(z)=\int_{0}^{z} {\rm d}w\,
(1+w)^{-2} \left[ \Omega_0(1+w)^3+\Omega_{\Lambda}\right]^{-1/2}$.
For $\Omega_{\Lambda}=0$,
$\tau_L^p(z) = \frac{3}{5}\, \Omega_{\rm L}\, \left[ (
y^{5/2}+1) \ln y /(y^{5/2}-1)   -\frac{4}{5} \right]$,
where $y=1+z$.


\section{Predictions/Speculation}

To make more specific predictions we must choose parameters and to do this we
are forced to speculate on the rate of inspiral events and the sensitivity of
future gravitational wave detectors.  Typically we expect neutron star 
binaries to have ${\cal M}_0=1.2\,{\rm M}_{\odot}$ while the advanced 
LIGO might have $r_0=355\,$Mpc, 
$\rho\geq \rho_0=5$ then implies $A=0.7573$.  We parametrize
evolutionary effects by $\eta(z)=(1+z)^{\beta}$, with $\beta\geq 0$ and a
redshift cut-off of $z_{\rm stop}=2.5$. The typical lifetime of a neutron star
binary is about $10^8$ to $10^9$ years \cite{Phinney91}; a significant fraction
of neutron star binaries formed at $z=3$ would have merged by $z=2.5$. Since
there seem to be a lot of star formation at $z>3$ \cite{Steidel96}, $z_{\rm
stop}=2.5$ is probably reasonable.

We find that the macrolensing rate is largest when there is a sizable
cosmological constant, but is still negligible unless there is significant
evolution.  Even with extreme evolution, e.g. $\Omega_{\Lambda}=0.8$,
$\Omega_0=0.2$, $F=0.05$ \cite{FT91}
and $\beta=3$ macrolensing yields only ~70\% of the
number of lensed events of microlensing with more modest parameters:
$\Omega_{\rm L}=0.07$ and $\beta=0$. Macrolensing of gravitational waves due to
galaxies is negligible compared to a plausible microlensing rate.  This is
partly due to the fact that we have a good idea of the number and properties
of galaxies while we are more free to speculate on the number of compact
objects and partly due to the fact that point mass lenses are more effective
gravitational lenses than galaxies.

For the rest of the paper we restrict ourselves to microlensing, which gives 
a few strongly lensed events per year without much evolution, 
for a currently acceptable
value of $\Omega_{\rm L}$.  We have considered two plausible cosmological
models: (1) $\Omega_0=1$ and $\Omega_{\rm L}=0.1$; (2) $\Omega_{\Lambda}=0.8$,
$\Omega_0=0.2$, $\Omega_{\rm L}=0.07$.

We consider expected data with cut on the signal-to-noise ratio, $\rho >5$.
Fig.1 shows the expected total number per year of events 
which are present due to gravitational lensing as function of
$h$, for two cosmological models, with $\beta=0,1$.  The solid lines are for
$\Omega_0=1$ and $\Omega_{\rm L}=0.1$, and the dashed lines are for
$\Omega_{\Lambda}=0.8$, $\Omega_0=0.2$, and $\Omega_{\rm L}=0.07$.  
Fig.2 shows the corresponding expected total number per year
of events which can be seen without gravitational lensing as function
of $h$, with the same line types as Fig.1.  The expected total numbers 
in both Fig.1 and Fig.2 increase with increasing $\beta$, as expected.
Fig.3 shows the redshift distribution of expected events corresponding to 
Figs.1-2 for $h=0.8$.  The dotted lines indicate the distribution 
of expected events which are present due to gravitational lensing.  
Note that gravitational lensing leads to tails at high redshift.  
For each cosmological model, the higher tail corresponds to $\beta=1$.  
Note that in principle, the evolutionary index can be measured from 
the region of the redshift distribution dominated by events which can
be seen without gravitational lensing.  Note also that most of the
events which are seen due to gravitational lensing lie beyond the 
cut-off redshift of the events which can be seen without gravitational
lensing.

We have used ${\cal M}_0=1.2 \,M_{\odot}$ as the typical intrinsic chirp
mass of neutron star binary inspirals.  It is expected that ${\cal M}_0$ will
fall in the narrow  range of $1.12 -1.26\,M_{\odot}$\cite{Finn96} while an
advanced LIGO can  measure the observed chirp mass ${\cal M}=(1+z) {\cal M}_0$
to an accuracy of better than 0.1\% \cite{FinnCher93,CutFla94}.  Thus the
uncertainty in the redshift of a given event will be very small compared to the
large range of $z$ over which the events which are seen due to
gravitational lensing are distributed [see Fig.3].
Since the redshift distribution of observed events which can be seen
without gravitational lensing should terminate at
a relatively small redshift $z_{\rm max}$, an observation of an event with
redshift significantly greater than $z_{\rm max}$ is a strong evidence for
gravitational lensing.  One should be able to identify the events
which are seen due to gravitational lensing!

\section{Discussion}

While most neutron star binary inspiral events detected by an advanced 
LIGO will probably not be affected by gravitational lensing, there could 
be a detectable number of events
which are  significantly magnified via gravitational lensing by compact
objects. These lensed events will be easily identifiable by their high
observed chirp masses. For the no-evolution parameters used above one would
expect around two events per year which are seen due to gravitational
lensing. Even a modest evolution ($\beta=1$)
of the rate of inspirals can significantly increase the rate of events
which are seen due to gravitational lensing,
and one could imagine even stronger evolution. During the lifetime of a
detector, say ten years, one might detect dozens of events
which are seen due to gravitational lensing, from which
one could estimate the amount of matter in compact lenses, $\Omega_{\rm L}$.
The absence of such expected lensed events will place an interesting constraint
on $\Omega_{\rm L}$. 

If we can determine $\Omega_{\rm L}$ in this way, we will have a much better
handle on the nature of the dark matter in our Universe.  Thus lensing adds
utility to the observation of inspiral events, which has already been 
shown to provide a measure of the Hubble constant, 
the deceleration parameter, and the cosmological
constant \cite{CherFinn93,Markovic93,Finn96,Wang96}. Gravitational lensing will
also add additional noise to the determination of these cosmological
parameters, although this noise is relatively small \cite{Markovic93}.  This is
because, as we have seen, most inspiral events are little affected by
gravitational lensing.  Finally we note that our consideration of lensing for
inspiral events is much the same as that which one uses when considering 
supernovae ``standard candles'' \cite{Linder88,Josh96}.

\acknowledgments
Y.W. and A.S. are supported by the DOE and NASA under Grant NAG5-2788.
E.L.T. gratefully acknowledges support from NSF grant AST94-19400.  We thank
Josh Frieman for very helpful discussions.


\narrowtext

\newpage
\nonfrenchspacing
\parindent=20pt
\centerline{{\bf Figure Captions}}

Fig.1 The total number per year of expected events which are present
due to gravitational lensing as function of $h$, with $\beta=0,1$.
The solid lines are for $\Omega_0=1$ and $\Omega_{\rm L}=0.1$,
and the dashed lines are for $\Omega_{\Lambda}=0.8$, $\Omega_0=0.2$,
and $\Omega_{\rm L}=0.07$.

Fig.2 The total number per year of expected events which can be seen
without gravitational lensing as function of $h$, 
with the same line types as Fig.1.

Fig.3 The redshift distribution of expected events
corresponding to Fig.1 for $h=0.8$.
The dotted lines indicate the distribution of expected events
which are seen due to gravitational lensing.
For each cosmological model, the higher tail corresponds to $\beta=1$.

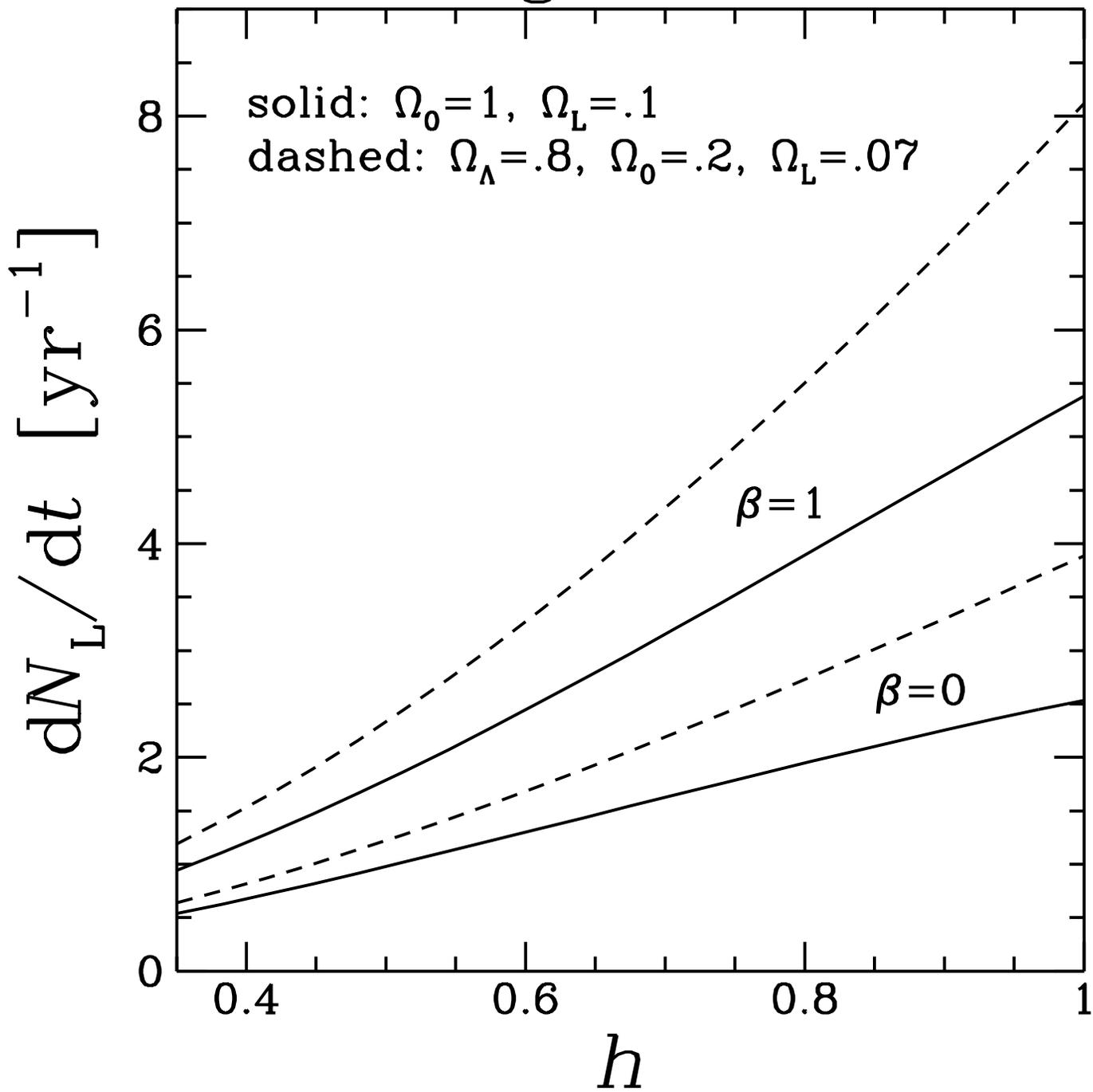
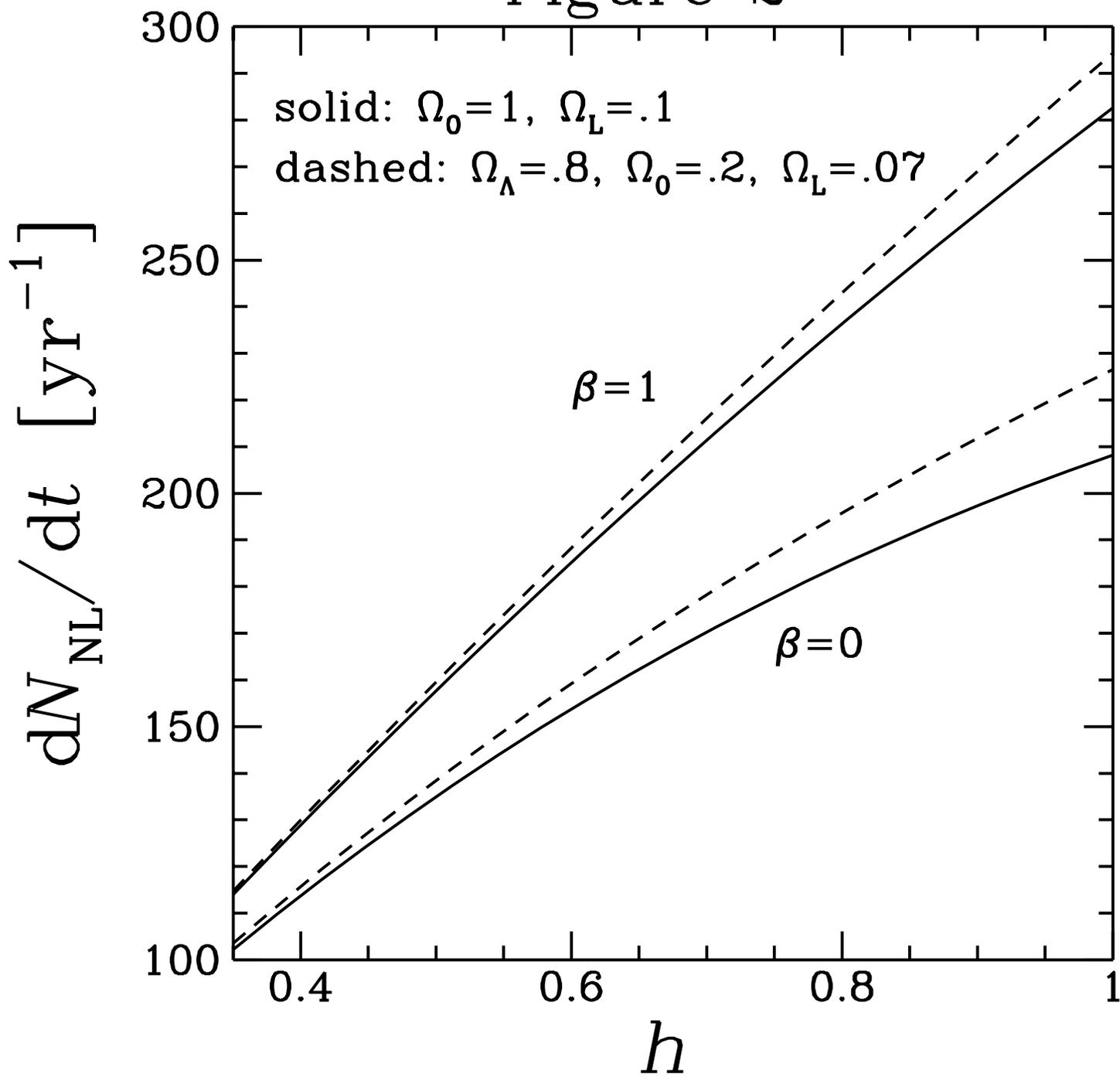
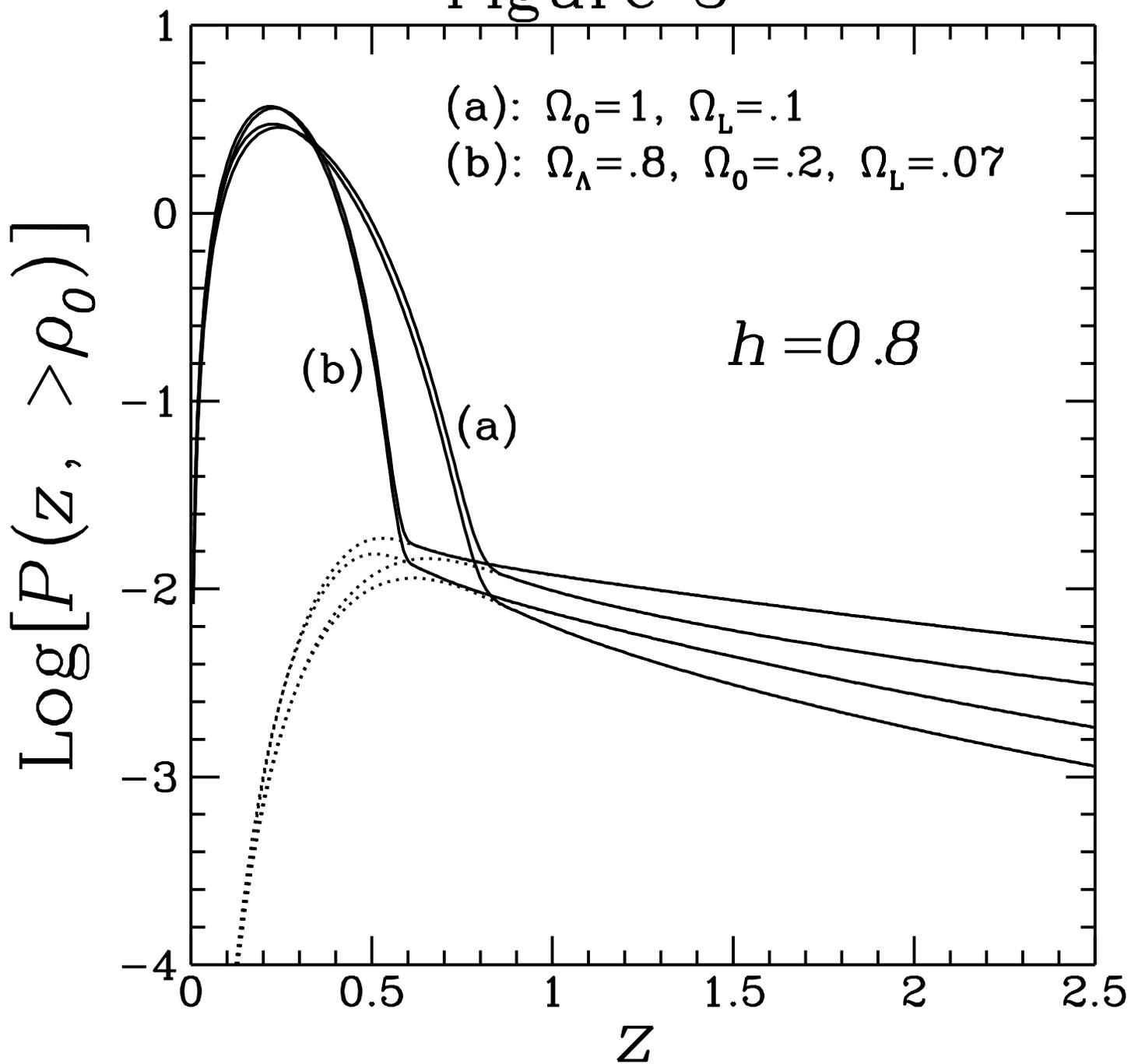
\end{document}